\documentclass[aps,prb,preprint,showpacs,showkeys,preprintnumbers,amsmath,amssymb]{revtex4}

\usepackage{graphicx}
\usepackage{dcolumn}
\usepackage{bm}

\begin{document}


\title{Cell size distribution in a random tessellation of space governed by the Kolmogorov-Johnson-Mehl-Avrami model: grain size distribution in crystallization}

\author{Jordi Farjas}
  \email{jordi.farjas@udg.cat}
  \homepage{http://copernic.udg.es/JordiFarjas/jfarjas.htm}
\author{Pere Roura}
  \email{pere.roura@udg.cat}
\affiliation{GRMT, Department of Physics,
	University of Girona, Campus Montilivi,
	Edif. PII, E17071 Girona, Catalonia, Spain}

\date{\today}

\begin{abstract}
The space subdivision in cells resulting from a process of random nucleation and growth is a subject of interest in many scientific fields. In this paper, we deduce the expected value and variance of these distributions while assuming that the space subdivision process is in accordance with the premises of the Kolmogorov-Johnson-Mehl-Avrami model. We have not imposed restrictions on the time dependency of nucleation and growth rates. We have also developed an approximate analytical cell-size probability density function. Finally, we have applied our approach to the distributions resulting from solid phase crystallization under isochronal heating conditions.
\end{abstract}


\maketitle

\section{INTRODUCTION}

In this paper we consider the subdivision of a $D$-dimensional Euclidean space into disjoint regions created after a process of random nucleation and growth. Random subdivisions can be obtained by several different methods, amongst which  Poisson-Voronoi and Johnson-Mehl tessellations \cite{Gilbert1962} have been widely studied. The Poisson-Voronoi tessellation is obtained by randomly picking several points, the seeds $P_{i}$, by a Poisson process. Next, the space is subdivided in cells, $C_{i}$, by the rule: $C_{i}$ contains all points in space closer to $P_{i}$ than to any other seed. This cellular structure is extensively applied in many diverse scientific fields including biology, \cite{Hamilton1971,HONDA1978} computer science, \cite{Haralick1979,Aurenhammer1991} materials science, \cite{Srolovitz1986,DiCenzo1989} astrophysics, \cite{Martinez1990,Scherrer1986,Yoshioka1989} medicine, \cite{Moroni2008} agriculture, \cite{Fischer1973} quantum field theory, \cite{Drouffe1984} and sociology. \cite{Boots1975}

The space tessellation can be fully characterized by means of the probability density function (PDF), $f(s)$, which is the probability that a cell has a size between $s$ and $s+ds$. The properties of the PDF of the Poisson-Voronoi tessellation have been extensively studied both theoretically \cite{Gilbert1962,Meijering1953,Pineda2004} and numerically. \cite{DiCenzo1989,Kiang1966,Weaire1986,Tanemura2001,Hinde1980,Kumar1992,Pineda2004,Ferenc2007} It is well known that the Poisson-Voronoi tessellation PDF is described by a gamma distribution

\begin{equation}
f(s)=\left( \frac{\nu}{E} \right)^{\nu}
\frac{1}{\Gamma(\nu)}s^{\nu-1}
\text{exp}\left(-\frac{\nu}{E}\, s\right)
\label{eq:gammadist}
\end{equation}

\noindent where $\Gamma$ is the gamma function, $\nu$ is a parameter that is dependent on the dimension $D$, i.e. $\nu=2$, 3.584 and 5.586 for $D=1$, 2 and 3 respectively, and $E$ is the expected cell size,

\begin{equation}
E \equiv \int_0^{\infty} s f(s) ds
\label{eq:Egammadist}
\end{equation}

It is worth mentioning that Eq.~(\ref{eq:gammadist}) has been analytically derived for the one-dimensional case where $\nu$=2 is an exact result. \cite{Meijering1953} Conversely, for the two and three-dimensional cases, the validity of Eq.~(\ref{eq:gammadist}) is supported by analytical approximations and numerical fits.

Our main interest is the characterization of grain morphology related to crystallization. In general, the crystallization of most materials takes place by means of a nucleation and growth mechanism: nucleation starts with the formation of small atom clusters of the new stable phase in the metastable phase. Subsequently, clusters with sizes greater than the critical, or nuclei, start to grow by incorporating neighboring atoms of the metastable phase. During this growth, grains impinge upon each other. Finally, the structure of the new stable phase consists of disjoint regions or crystals separated by grain boundaries. The evolution of crystallization and grain size distributions is entirely determined by the nucleation rate density $I$ and the grain linear growth rate $G$. When nucleation takes place for a very short time, its rate may vanish before the onset of particle growth (\textit{site saturated nucleation}). \cite{Cahn1956,Frade1998} In this case, the crystal structure is equivalent to a Poisson-Voronoi tessellation provided that nucleation is Poissonian through the whole space and growth is isotropic.

Conversely, \textit{continuous nucleation} takes place when nucleation and growth occur at the same time. In general, there is an energy barrier for nucleation and growth to happen. Thus, $I$ and $G$ depend on temperature. For the particular case of isotropic and isothermal transformations, where $I$ and $G$ are constant, the resulting crystal structure corresponds to the well-known Johnson-Mehl tessellation. \cite{Gilbert1962} For this tessellation, Axe et al. \cite{Axe1986} have obtained an analytical solution for the one-dimensional case while Mulheran \cite{Mulheran1991} has developed a simple (but not so accurate) relation for the two- and three-dimensional cases. Alternatively, Monte-Carlo simulations provide a powerful tool for the calculation of tessellations and PDFs under a wide variety of conditions. \cite{Srolovitz1986,Rollett1997,Zhu1992,Crespo1996,Almansour1996,Pusztai1998,Castro1999,Kooi2004,Farjas2007,Farjas2008}

Under non-isothermal conditions, $I$ and $G$ depend on time by virtue of their temperature dependence. Therefore, an infinite number of different tessellations/structures can be obtained by varying the thermal history. Unlike the Poisson-Voronoi and Johnson-Mehl tessellations, the analytical results related to tessellations emerging from time dependent nucleation and growth rates are scarce. Indeed, as far as we know, the analytical models are limited to time dependent nucleation rates. \cite{Pineda2004,Jun2005} In particular, Jun et al. \cite{Jun2005} have derived an analytical solution for the one-dimensional case. Particularly relevant to the present work are the results of Pineda et al.\cite{Pineda2004} who have obtained an accurate analytical description for the two- and three-dimensional cases.

In the present work we will consider those transformations that fulfill the Kolmogorov-Johnson-Mehl-Avrami (KJMA) premises. No restrictions will be imposed on nucleation and growth rate time dependence. We will refer to these tessellations as \textit{KJMA tessellations}. KJMA theory has been widely applied to describe systems undergoing first-order phase transformations. For instance, DNA replication; \cite{Jun2005} crystallization of polymers, \cite{Long1995} amorphous materials \cite{Farjas2004,Farjas2008Aiche} and glasses; \cite{Henderson1979} switching in ferroelectrics \cite{Ishibash1971} and ferromagnets; \cite{Hirsch1992} lattice-gas models; \cite{Ramos1999} and film growth on solid substrates. \cite{Fanfoni2005} In Section II we will describe the basic concepts of KJMA theory and will focus our attention on those aspects that are useful to the development of our work. Section III is devoted to the calculation of the expected value and variance of the distributions related to the KJMA tessellations. In Section IV we will derive a simpler  approximate relation for the variance and will check its accuracy. As an application of the previous results, in Section V we will derive an approximate grain size PDF which is the superposition of gamma distributions. Finally, at the end this section we will verify that the grain radius PDF can be expressed as well as the superposition of Gaussian distributions.

\section{THE KOLMOGOROV-JOHNSON-MEHL-AVRAMI THEORY}

The KJMA theory \cite{Avrami1939,Johnson1939,Kolmogorov1937} describes in a very simple form the kinetics of transformations governed by nucleation and growth that satisfy the following assumptions:

\renewcommand{\labelenumi}{\roman{enumi}}

\begin{enumerate}
\item nucleation must be Poissonian through the entire space;
\item the volume of an arbitrary grain is much smaller that the volume of the system;
\item the crystal growth rate is isotropic.
\end{enumerate}

On the basis of these premises, Kolmogorov calculated the evolution of the transformed fraction, $X(t)$, through the probability, $p(t)$, that an arbitrary point $O$ has not crystallized, i.e.,  the probability that no nuclei able to transform $O$ will be formed during the time interval $[0,t]$,
\begin{subequations}
\label{eq:kolmogorov}
\begin{equation}
 X(t)=1-p(t) \label{subeq:kol1},
\end{equation}
\begin{equation}
 p(t)=\exp \left[-g_D\int_0^t I(\tau) r(t,\tau)^D d\tau \right]
\label{subeq:kol2},
\end{equation}
\begin{equation}
r(t,\tau) \equiv \int_{\tau}^t G(z) d z
\label{subeq:kol3},
\end{equation}
\end{subequations}

\noindent where $g_D$ is a geometrical factor related to the shape of the crystal -- for a $D$-dimensional sphere $g_D=\pi^{D/2} / \Gamma(D/2+1)$ -- and $r(t,\tau)$ is the minimum distance between $O$ and a nucleus created at $\tau$, so that the nucleus would not transform $O$.

Based on geometrical arguments, Avrami deduced the following relation:
\begin{equation}
\frac{\partial_t v(t,\tau)}{\partial_t v_{ex}(t,\tau)}=\frac{1-X(t)}{1-X(\tau)}
\label{eq:Avrami1},
\end{equation}
\noindent where $\partial_t v(t,\tau)$ and $\partial_t v_{ex}(t,\tau)$ are respectively the actual and \textit{extended} average volumetric growth rate at time $t$ for grains nucleated at time $\tau$. The word \textit{extended} refers to the volume a grain would attain if nuclei grew through each other and overlapped without mutual interference.

The integration of Eq.~(\ref{eq:Avrami1}) leads to\cite{Sessa1996}
\begin{equation}
\frac{dX(t)}{1-X(t)}=dX_{ex}(t)
\label{eq:Avrami2}.
\end{equation}
Finally, integration of Eq.~(\ref{eq:Avrami2}) gives Avrami's well-known formula
\begin{equation}
X(t)=1-\exp \left[-X_{ex}(t) \right]
\label{eq:Avrami3}.
\end{equation}
The calculation of $X_{ex}(t)$ is straightforward and obtained by simply neglecting the impingement between nuclei
\begin{equation}
 X_{ex}(t)=g_D \int_0^t I(\tau) r(t,\tau)^D d\tau
\label{eq:Xex}.
\end{equation}
The combination of Eqs.~(\ref{eq:Avrami3}) and (\ref{eq:Xex}) gives Eq.~(\ref{eq:kolmogorov}). As it is well known, Avrami and Kolmogorov deduced the same relation using different approaches.

Note that in Eq.~(\ref{eq:Xex}) it is assumed that the nucleation rate is not affected by the shrinking of the untransformed phase. In the calculation of $X_{ex}(t)$ the \textit{phantom nuclei} are taken into account. Avrami designated as phantom nuclei those nuclei that are formed in the transformed fraction and therefore do not contribute to the formation of new grains. Indeed, the actual nucleation rate can be defined as
\begin{equation}
 I_a(t) \equiv \left[ 1-X(t) \right] I(t)
\label{eq:Iact}.
\end{equation}

Concerning the limitations of the KJMA theory, it also holds in the case of anisotropic growth provided that the grains have a convex shape and are aligned in parallel.\cite{Pusztai1998} Moreover, the KJMA theory provides a good approximation when the anisotropy is moderate or for soft impingement.\cite{Bruna2006} However, KJMA theory fails when nucleation is non-random, \cite{Tomellini2008} when growth is anisotropic, \cite{Birnie1995,Kooi2004,Burbelko2005} when growth stops before crystallization is complete\cite{Bruna2006} and when the incubation time is not negligible.\cite{Tomellini1997}

\section{Statistical properties of the KJMA cell size distribution}

The cell size distribution is characterized by its PDF, $f(s)$, the probability that a cell has a size between $s$ and $s+ds$. From its definition it is obvious that $f(s)$ must be normalized
\begin{equation}
\int_{0}^{\infty} f(s) ds =1
\label{eq:PDFNorm}.
\end{equation}
To analyze the properties of the cell size distribution, we will consider the contribution of the crystals formed at a time $\tau$ over a time interval $d\tau$ ($\tau$-crystals). We will call the cell size distribution of the $\tau$-crystals the $\tau$-distribution. Accordingly, we define the PDF of the $\tau$-crystals, $f_{\tau}(s)$, as the probability that a $\tau$-crystal has a volume between $s$ and $s+ds$. From the definition of $f_{\tau}(s)$, it is also apparent that $f_{\tau}(s)$ must be normalized:
\begin{equation}
\int_{0}^{\infty} f_{\tau}(s) ds =1
\label{eq:PDFparcial}.
\end{equation}
$f(s)$ is simply the addition of the contributions of the $\tau$-crystals over the time interval in which their nucleation takes place
\begin{equation}
 f(s)=\frac{\int_0^{\infty}I_a(\tau) f_{\tau}(s) d \tau}{\int_0^{\infty}I_a(\tau) d \tau}
\label{eq:PDFtotal}.
\end{equation}
Note that the denominator in Eq.~(\ref{eq:PDFtotal}) ensures that $f(s)$ is normalized if all $f_{\tau}(s)$ are normalized.

In the following sections, we will present the expected grain size and the variance of the cell size distribution and their relationship with the equivalent parameters of the $\tau$-distributions.

\subsection{Expected grain size}

It is well known that the expected grain size, $E$, is the inverse of the final grain density:
\begin{equation}
E=\left( \int_0^{\infty} I_a(\tau) d \tau \right)^{-1}
\label{eq:EsExt}.
\end{equation}

Likewise, the expected value, $E_{\tau}$ of a $\tau-$distribution is simply the final average grain size of a $\tau$-crystal normalized to the total volume:
\begin{equation}
E_{\tau}= \int_{\tau}^{\infty}\partial_z v(z,\tau) d z
\label{eq:Etau1}.
\end{equation}
Introducing Eq.~(\ref{eq:Avrami1}) into Eq.~(\ref{eq:Etau1}) leads to
\begin{equation}
E_{\tau}=\frac{1}{1-X(\tau)}\int_{\tau}^{\infty} \left[ 1-X(z) \right] \, \partial_z v_{ex}(z,\tau) d z
\label{eq:Etau2},
\end{equation}
where the extended average growth rate is given by
\begin{equation}
 \partial_z v_{ex}(z,\tau)= D g_D \, r (z,\tau) ^{D-1} G(z)
\label{eq:vex}.
\end{equation}

Note than once the evolution of the transformed fraction, $X(t)$, is known -- i.e. the solution of Eq.~(\ref{eq:kolmogorov}) --  the calculation of $E_{\tau}$ is straightforward.

Besides, the final space fraction occupied by the $\tau$-crystals, $X_{\tau}$, can be calculated from the integration over the entire space of the probability that a point $P$ in the space belongs to a tau crystal nucleated at $O$.
Since the system is homogeneous and isotropic, this probability only depends on the distance $b$ between $O$ and $P$. Therefore,
\begin{eqnarray}
 X_{\tau} & = & I (\tau) \! \int \! P_{\tau}(O,P) d V_P = \nonumber \\
          & = & D g_D I (\tau) \int_0^\infty \!\! P_{\tau}(b) b^{D-1} d b
\label{eq:Xtau1},
\end{eqnarray}

\noindent where $P_{\tau}(b)$ is the probability that a point $P$, separated by a distance $b$ from the nucleus $O$, belongs to the crystal nucleated at $O$. To calculate $P_{\tau}(b)$, we will use the same approach that Kolmogorov used for the deduction of Eq.~(\ref{eq:kolmogorov}). Since nucleation is Poissonian, $P_{\tau}(b)$ is given by the probability that no nucleus is formed that could transform $P$ before $O$ does so. $P$ would be transformed by $O$ at the moment $t_b$:
\begin{equation}
b = r(t_b,\tau)=\int_{\tau}^{t_b} G(z) d z
\label{subeq:tb}.
\end{equation}

\noindent Thus, the nuclei formed at $z$ that could transform $P$ before $O$ does so, are located in a $D$-sphere of radius $r(t_b,z)$ around $P$. Therefore, according to Eq.~(\ref{eq:kolmogorov}), $P_{\tau}(b)$ is given by
\begin{equation}
P_{\tau}(b)=\exp \left[ - g_D \int_0^{t_b} I(z) r(t_b,z) ^D dz \right]
\label{eq:Ptaub1}.
\end{equation}

\noindent The previous integral spans the time interval $[0,t_b]$ since no nucleus formed after $t_b$ could transform $P$. Comparison of Eq.~(\ref{eq:Xex}) with Eq.~(\ref{eq:Ptaub1}) gives
\begin{equation}
P_{\tau}(b)=\exp \left[ - X_{ex}(t_b)\right] =1-X(t_b)
\label{eq:Ptaub2}.
\end{equation}

Finally, if we introduce the value of $P_{\tau}(b)$ given by Eq.~(\ref{eq:Ptaub1}) into Eq.~(\ref{eq:Xtau1}) and we change the variable $b$ by $t_b$, we obtain
\begin{equation}
X_{\tau} \! = \! D g_D I(\tau) \! \int_{\tau}^{\infty} \! \left[ 1-X(t_b) \right] r (t_b,\tau) ^{D-1} G(t_b) d t_b
\label{eq:Xtau2}.
\end{equation}

Alternatively, the expected value, $E_{\tau}$ is the ratio between the space fraction occupied by the $\tau$-crystals and the density of $\tau$-crystals:
\begin{equation}
E_{\tau}=\frac{X_{\tau} d \tau}{I_a({\tau}) d \tau}
\label{eq:Etau3}.
\end{equation}

As expected, substitution of Eqs.~(\ref{eq:Xtau2}) and (\ref{eq:Iact}) into Eq.~(\ref{eq:Etau3}) delivers Eq.~(\ref{eq:Etau2}). Moreover, the integration of $X_{\tau}$ over the whole time interval where nucleation takes place gives the total transformed fraction, $1$:
\begin{equation}
\int_0^{\infty} I_a(\tau) E_{\tau} d \tau = \int_0^{\infty} X_{\tau} d \tau = 1
\label{eq:IntegXtau}.
\end{equation}

We will end this subsection verifying that the value of $E$ evaluated from the $\tau$-distributions coincides with
the value given at the beginning of this subsection [Eq.~(\ref{eq:EsExt})]:

\begin{eqnarray}
E\!& \equiv & \! \int_{0}^{\infty} s f(s) ds =
	\frac{\int_0^{\infty} I_a(\tau) \left( \int_0^{\infty} s f_{\tau}(s)ds \right) d \tau}
	 {\int_0^{\infty} I_a(\tau) d \tau}=\nonumber \\
    & =& \frac{\int_0^{\infty} I_a(\tau) E_{\tau}d \tau} {\int_0^{\infty} I_a(\tau) d \tau}=
	\frac{1} {\int_0^{\infty} I_a(\tau) d \tau}
 \label{eq:EsPDF}.
\end{eqnarray}

\subsection{Variance of the grain size distribution}

To determine the variance we will adapt the development of Gilbert\cite{Gilbert1962} for a Poisson-Voronoi tessellation to our case. First, we define a new PDF, $f^*(s)$, as the PDF of the crystals that contain a given arbitrary point $O$: i.e., if we pick an arbitrary point $O$, $f^*(s)$ is the probability that a crystal has a size between $s$ and $s+ds$ and contains the point $O$. Accordingly, $f^*(s)$ is proportional to $f(s)$ and to $s$, because a large crystal has a proportionally greater chance of containing the point $O$. Therefore,
\begin{equation}
f^*(s) = \frac{s \,f(s)}{E}
\label{eq:festrella}.
\end{equation}

The constant of proportionality, $E^{-1}$, has been deduced by imposing normalization:
\begin{equation}
\int_0^{\infty} f^*(s) d s = 1
\label{eq:normfestrella}.
\end{equation}

From the definition of $f^*(s)$ it can be easily proved that
\begin{equation}
 \mathrm{var}=\int_0^\infty (s-E)^2 f(s) d s = E^* E-E^2
\label{eq:variantotal}
\end{equation}
\noindent where $E^*$ is the expected value of $f^*(s)$.

Once $E^*$ is known, the calculation of the variance is simple. To obtain $E^*$ we first analyze the contribution of the $\tau$-crystals. To do so, we define $f_{\tau}^*(s)$ as the PDF of the $\tau$-crystals that contain a given arbitrary point $O$: i.e., if we pick an arbitrary point $O$, $f_{\tau}^*(s)$ is the probability that a $\tau$-crystal has a volume between $s$ and $s+ds$ and contains the point $O$. Accordingly, $f_{\tau}^*(s)$ is proportional to $f_{\tau}(s)$ and to $s$:
\begin{equation}
f_{\tau}^*(s) \propto f_{\tau}(s) \, s
\label{eq:festtauprop}.
\end{equation}

On the other hand, the integration of $f_{\tau}^*(s)$ over all possible volumes is the probability that an arbitrary point $O$ belongs to a $\tau$-crystal. This probability is the fraction of the space occupied by the $\tau$-crystals $X_{\tau}$:
\begin{equation}
\int_0^{\infty} f_{\tau}^*(s) d s = X_{\tau}
\label{eq:festtaunorm},
\end{equation}
\noindent taking into account that,
\begin{equation}
E_{\tau}=\int_0^{\infty} s f_{\tau}(s) d s
\label{eq:defEtau},
\end{equation}
\noindent and combining Eqs.~(\ref{eq:festtauprop}), (\ref{eq:festtaunorm}) and (\ref{eq:Etau3}) we obtain $f_{\tau}^*(s)$:
\begin{equation}
f_{\tau}^*(s)= I_a(\tau) \, s \, f_{\tau}(s)
\label{eq:PDFestrella}.
\end{equation}

Then the expected value of $f_{\tau}^*$, $E_{\tau}^*$, is 
\begin{equation}
E_{\tau}^* = \frac{\int_0^{\infty} s f_{\tau}^*(s) d s}{\int_0^{\infty} f_{\tau}^*(s) d s}
\label{eq:Etauestrella}.
\end{equation}

It can be easily verified that $E^*$ is related with $E_{\tau}^*$ through
\begin{equation}
E^* = \frac{\int_0^{\infty} s f^*(s) d s}{\int_0^{\infty} f^*(s) d s}=\int_0^{\infty} X_{\tau} E_{\tau}^* d{\tau}
\label{eq:Eestrella}.
\end{equation}

Therefore the contribution of the $\tau$-crystal to the expected value $E^*$ is $X_{\tau} E_{\tau}^*$. In addition, this contribution is the integration over the entire space of the probability that a differential volume around a point $P$ in the space belongs to the same $\tau$-crystal as $O$. Since the system is homogeneous and isotropic, this probability only depends on the distance $b$ between $O$ and $P$,
\begin{equation}
X_{\tau} E_{\tau}^*  = D \, g_D  \int_0^\infty  P_{\tau}^*(b) b^{D-1} d b
\label{eq:Eestrellatau},
\end{equation}

\noindent where $P_{\tau}^*(b)$ is the probability that two points, $O$ and $P$, separated by a distance $b$ belong to the same $\tau$-crystal,
\begin{equation}
P_{\tau}^*(b)=I(\tau) \int P_{\tau}^*(b,Q) d V_Q
\label{eq:PbEst}.
\end{equation}

\begin{figure}
\includegraphics[width=8.3cm]{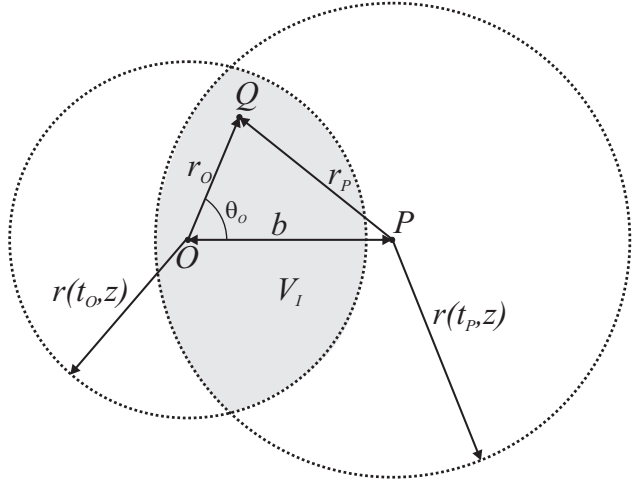}
\caption{\label{fig:esquema} Schematic drawing of the calculation of $P_{\tau}^*(b,Q)$.}
\end{figure}

The integration domain covers the entire space, $d V_Q$ is the $D$-volume differential around a point $Q$, $I(\tau) d V_Q d \tau$ is the probability that a nucleus is formed at $Q$ at the time $\tau$ during the time interval $d \tau$ and $P_{\tau}^*(b,Q)$ is the probability that both $O$ and $P$ belong to the same crystal nucleated at $Q$ (see Fig.~\ref{fig:esquema}). For $D=2$, $d V_Q=2 r_O d r_O d \theta_O$ while for $D=3$, $d V_Q=2 \pi r_O^2 \sin(\theta_O) d r_O d \theta_O$ in polar coordinates.

$P_{\tau}^*(b,Q)$ is given by the probability that no nucleus is formed that could transform $O$ or $P$ before $Q$ does, then
\begin{subequations}
\label{eq:PbQ}
\begin{eqnarray}
\lefteqn{ P_{\tau}^*(b,Q)\!=\! \exp \Big\{ -g_D \Big[ \int_0^{t_O} I(z) r(t_O,z)^D dz \, + }\nonumber \\
    & &\! +\! \int_0^{t_P} \! I(z) r(t_P,z)^D dz  - \int_0^{t'} \! I(z) \frac{V_I}{g_D} dz  \Big] \Big\}
 \label{subeq:Pb1},
\end{eqnarray}
\begin{equation}
r_P^2= r_O^2+b^2-2 r_O b \cos \theta_O
\label{subeq:Pb2},
\end{equation}
\begin{equation}
r_x = r(t_x,\tau)=\int_{\tau}^{t_x} G(z) d z, \quad \mathrm{for}\ \, x=O, P
\label{subeq:Pb3}
\end{equation}
\begin{equation}
\frac{r_O+r_P-b}{2} = r(t',\tau)=\int_{\tau}^{t'} G(z) d z
\label{subeq:Pb4},
\end{equation}
\end{subequations}

\noindent where $r(t_O,z)$ and $r(t_P,z)$ are the minimum distance between $O$, $P$ and a nucleus created at the time $z$, so that the nucleus would not transform $O$ and $P$ respectively (see Fig.~\ref{fig:esquema}). $V_I$ is the volume intersection between two $D$-spheres of radius $r(t_O,z)$ and $r(t_P,z)$ centered at $O$ and $P$, respectively (gray region in Fig.~\ref{fig:esquema}). The subtraction of the term $V_I$ is in accordance with the fact that it has been accounted twice in the first and second integrals in Eq.~(\ref{subeq:Pb1}). For a particular set of values of the integration variables $r_O$ and $\theta_O$, $r_P$ is evaluated from Eq.~(\ref{subeq:Pb2}) while $t_O$ and $t_P$ are defined by Eq.~(\ref{subeq:Pb3}) and $t'$ is defined by Eq.~(\ref{subeq:Pb4}). Note that $O$ and $P$ are transformed by $Q$ at the times $t_O$ and $t_P$, respectively. Thus, any nucleus formed after $t_O$ and $t_P$ could not transform $O$ or $P$, respectively, so the two first integrals in Eq.~(\ref{subeq:Pb1}) span the time interval $[0,t_O]$ and $[0,t_P]$, respectively. Additionally, it can be easily verified that if $z>t'$, then the intersection between the $D$-spheres is null. Therefore, the last integral in Eq.~(\ref{subeq:Pb1}) spans the time interval $[0,t']$.

Finally, Eq.~(\ref{eq:PbQ}) is simplified by substitution of Eq.~(\ref{eq:Xex}) in the first and second integrals in Eq.~(\ref{subeq:Pb1})
\begin{eqnarray}
P_{\tau}^*(b,Q)\!& = & \! \frac{\exp \left[ - X_{ex}(t_O) - X_{ex}(t_P)  \right]}
				 {\exp \left[ - \int_0^{t'} I(z) V_I dz \right]} \nonumber \\
    & = &  \frac{ \big[ 1 - X(t_O) \big] \big[ 1 - X(t_P) \big]} {\exp \left[ - \int_0^{t'} I(z) V_I dz \right]}
 \label{eq:PbQfinal}.
\end{eqnarray}

It can be easily proved that the variance of the $\tau$-distributions, $\mathrm{var}_{\tau}$, is given by
\begin{equation}
 \mathrm{var}_{\tau} = E_{\tau}^*E_{\tau} -(E_{\tau})^2
\label{eq:vartau}.
\end{equation}

Finally, we will check if the variance of the distribution determined from the decomposition of $f(s)$ into $\tau$-PDF, Eq.~(\ref{eq:PDFtotal}), gives the expected result, Eq.~(\ref{eq:variantotal}):
\begin{eqnarray}
\mathrm{var} & = & \frac{\int_0^{\infty} \left( \int_0^{\infty} s^2  I_a(\tau) f_{\tau}(s)ds \right) d \tau}
	 {\int_0^{\infty} I_a(\tau) d \tau}-E^2=\nonumber \\
    & =& E \left[ \int_0^{\infty} \left( \int_0^{\infty} s  f_{\tau}^*(s)ds \right) d \tau \right]  -E^2=\nonumber \\
	& =& E \left[ \int_0^{\infty} E_{\tau} d \tau \right]  -E^2 = E^* E-E^2
 \label{eq:verifvar}.
\end{eqnarray}
At this point, we would like to point out that the results obtained so far are exact and general, i.e., we have not made any assumption concerning $f(s)$ and $f_{\tau}(s)$.

\section{Approximate variance}

According to our previous analysis, the exact calculation of the variance is reduced to the calculation of the parameters $E$ and $E^*$ in Eq.~(\ref{eq:variantotal}). While the calculation of $E$ is straightforward, the evaluation $E^*$ is more cumbersome. Indeed, when compared to Monte-Carlo algorithms,\cite{Farjas2007} its numerical calculation is more complex without representing any significant reduction in computing time. That is because there are several integrals nested and, in particular, the calculation of the intersection volume $V_I$ is complex. When $r(t_O,z)\gg b$ or $b\gg r(t_O,z)$, $V_I$ tends towards being a $D$-sphere of radius $r(t_O,z)$ and $0$, respectively. On the other hand, when $r(t_O,z)\approx r(t_P,z)$ the shape of $V_I$ roughly approaches a $D$-sphere. Since the width of $V_I$ (see Fig.~\ref{fig:esquema}) is $r(t_O,z)+r(t_P,z)-b$, we approximate $V_I$ by a $D$-sphere of diameter $r(t_O,z)+r(t_P,z)-b$:
\begin{equation}
V_I\approx g_D \left( \frac{r(t_O,z)+r(t_P,z)-b}{2} \right)^D
\label{eq:VIAprox}.
\end{equation}

It is worth noting that the previous approximation also works for the limiting cases $r(t_O,z)\gg b$ and $b\gg r(t_O,z)$. Furthermore, for the one-dimensional case it can be easily verified that Eq.~(\ref{eq:VIAprox}) is exact (in Appendix A we derive $P_{\tau}^*$ for $D=1$). Finally, the approximate solution (from here on \textit{approximation I)} is obtained by substitution of Eqs.~(\ref{eq:Xex}) and (\ref{eq:VIAprox}) into Eq.~(\ref{eq:PbQfinal}):
\begin{eqnarray}
P_{\tau}^*(b,Q)\!& = & \! \frac{\exp \left[ - X_{ex}(t_O) - X_{ex}(t_P)  \right]}
				 {\exp \left[ - X_{ex}(t') \right]} \nonumber \\
    & = &  \frac{ \big[ 1 - X(t_O) \big] \big[ 1 - X(t_P) \big]} {\big[ 1 - X(t') \big]}
 \label{eq:PbQaprox}.
\end{eqnarray}
Therefore, the calculation of $P_{\tau}^*(b,Q)$ is simple provided that the evolution of the transformed fraction, $X(t)$, is known. Analytical exact solutions for $X(t)$ are restricted to three particular situations under isothermal conditions: time-independent growth and nucleation rates, time-independent growth rate and nucleation rate proportional to a power of time,\cite{Ruitenberg2002} and site saturated nucleation. A quasi-exact solution of the KJMA model has recently been obtained under continuous heating conditions.\cite{Farjas2006} Moreover, there are numerical methods which allow a simple and fast calculation of $X(t)$ for an arbitrary time dependence of the nucleation and growth rates.\cite{Farjas2007}

We have analyzed the distribution emerging from solid phase crystallization under isochronal heating conditions, i.e. heating at a constant rate, to check the accuracy of \textit{approximation I}, Eq.~(\ref{eq:PbQaprox}), in the case of time dependent nucleation and growth rates. To work with realistic parameters we have taken those of amorphous silicon crystallization,\cite{Spinella1998,Farjas2004,Farjas2008Aiche} in which the nucleation and growth rates are described by an Arrhenius temperature dependence
\begin{equation}
 I\!=\!I_0 \exp \left( -\frac{E_N}{K_B T} \right) \ \textrm{and} \
 G\!=\!G_0 \exp \left( -\frac{E_G}{K_B T} \right)
\label{eq:Arrhenius}.
\end{equation}

\begin{table}
\caption{\label{tab:table1}Experimental nucleation and growth rates of amorphous silicon (Ref.~\onlinecite{Spinella1998,Farjas2004,Farjas2008Aiche}).}
\begin{ruledtabular}
\begin{tabular}{lll}
Nucleation & Activation energy, $E_N$ & 5.3 eV\\
 & Preexponential term, $I_0$ & $1.7\times 10^{44} \, \textrm{s}^{-1} \textrm{m}^{-3}$ \\
Growth & Activation energy, $E_G$  & 3.1 eV\\
 & Preexponential term, $G_0$ & $2.1 \times 10^7 \, \textrm{s}^{-1} \textrm{m}$\\
\end{tabular}
\end{ruledtabular}
\end{table}

\noindent where $T$ is the temperature in Kelvin and $k_B$ is the Boltzmann constant. In Table~\ref{tab:table1} we summarize the corresponding parameters. When the temperature is raised at a constant rate $\beta$, the nucleation and growth rates become time dependent through Eq.~(\ref{eq:Arrhenius}). Under those conditions, the kinetics is correctly described by the KJMA theory\cite{Farjas2006,Farjas2008} and there is good agreement between experiment and theoretical predictions.\cite{Farjas2004,Farjas2008Aiche} For the calculation of the evolution of the transformed fraction, we have used the quasi-exact solution described in Ref.~\onlinecite{Farjas2006}. The numerical evaluation of the integrals has been performed by means of an extended midpoint algorithm.\cite{Press2007} To confirm that the observed discrepancies are not related to numerical inaccuracies, we have performed several calculations with consecutive smaller integration steps. Moreover, for  the numerical integration over a semi-infinite interval, we have imposed a minimum relative error of $10^{-6}$. To check the accuracy of the numerical calculation, we have calculated the integral of $X_{\tau}$ over the interval $[0,\infty)$ and have compared them to its predicted value, Eq.~(\ref{eq:IntegXtau}). Calculations that exhibit discrepancies larger than $10^{-6}$ were rejected.

\begin{figure}
\includegraphics[width=8.3cm]{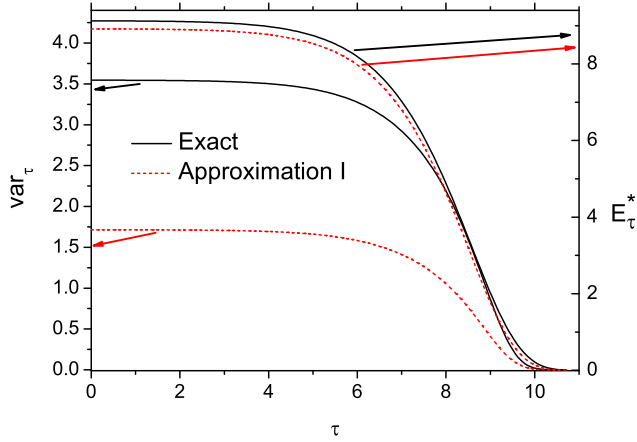}
\caption{\label{fig:vartau} $E_{\tau}^*$ and $\mathrm{var}_{\tau}$ for three-dimensional growth and isochronal heating of 40 K/min. Time and space have been normalized according to the time and space scaling factors, $(I(T_P)G(T_P)^3)^{-1/4}$ and $(G(T_P)/I(T_P))^{1/4}$, where $T_P$ is the peak temperature, i.e., the temperature at which the transformation rate is maximum (see Ref.~\onlinecite{Farjas2008}). The exact (black solid line) and the approximate (red dashed line) values are compared.}
\end{figure}

As is apparent from Fig.~\ref{fig:esquema}, the approximation of $V_I$ by a $D$-sphere of diameter equal to its width, Eq.~(\ref{eq:VIAprox}), results in an underestimation of $V_I$, which leads to an undervaluation of $E_{\tau}^*$ and of $\mathrm{var}_{\tau}$. The latter conclusion can be verified in Fig.~\ref{fig:vartau}, where the evolution of $E_{\tau}^*$ and $\mathrm{var}_{\tau}$ with $\tau$ is shown. Although \textit{approximation I} gives an accurate value of $E_{\tau}^*$, the approximate value of $\mathrm{var}_{\tau}$ shows a significant deviation from the exact value. The reason is that in the evaluation of $\mathrm{var}_{\tau}$, Eq.~(\ref{eq:vartau}), both terms in the difference have similar values. The same happens to the values of the variance and $E^*$; the exact and approximate values of $E^*$ are 3.93 and 3.69 respectively, while the exact and approximate variances are 3.56 and 3.22, respectively. (Space has been normalized  to the space scaling factor\cite{Farjas2008} $(G(T_P)/I(T_P))^{1/4}$, where $T_P$ is the peak temperature.) Despite the significant discrepancy between the exact and approximate values of $\mathrm{var}_{\tau}$, they have a nearly parallel evolution with $\tau$. This result is general and is related to the very similar dependency of the approximate and exact $V_I$ on the integration parameters. Therefore, the accuracy of \textit{approximation I} can be analyzed through the relation between the exact and approximate values of $\mathrm{var}_0$. To cover a wide range of distributions, we will recall the results given in Ref.~\onlinecite{Farjas2008}. In this work it was shown that the shape of the grain size PDF was practically insensitive to the heating rate, but it depends mainly on the ratio $E_N/E_G$, i.e., the relative evolution of the nucleation and growth rates with time. The limit $E_N/E_G \rightarrow 0$ corresponds to site saturated nucleation while $E_N/E_G=1$ coincides with the isothermal case.

\begin{figure}
\includegraphics[width=8.3cm]{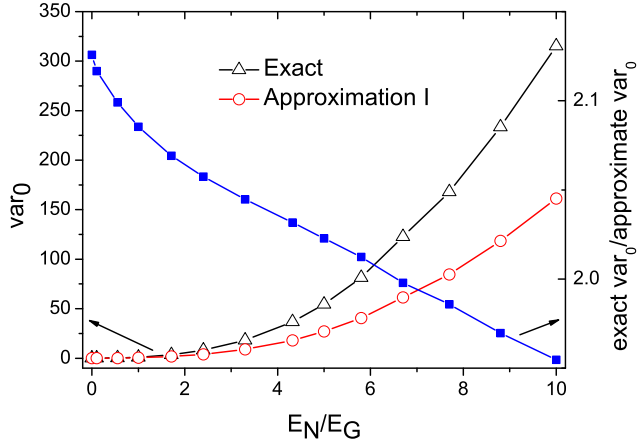}
\caption{\label{fig:evovartau} Exact (black triangles) and approximate (red circles) values of $\mathrm{var}_0$ and their ratio (solid blue squares) as a function of the ratio between nucleation and growth activation energies.}
\end{figure}

In Fig.~\ref{fig:evovartau} we have plotted the exact and approximate values of $\mathrm{var}_0$ as well as their ratio. (At this point it is worth recalling that, according to Eq.~(\ref{eq:Arrhenius}), a relation of one order of magnitude between the activation energies $E_N$ and $E_G$ would result in a huge difference in the relative time evolution between the nucleation and growth rates.) First, we can easily verify that $\mathrm{var}_0$ (and in general the variance) decreases with $E_N/E_G$. This means that the distributions become broader as $E_N/E_G$ increases. Indeed, when $E_G \gg E_N$, during the first stages of the transformation, nucleation dominates over growth. Most of the nuclei are formed at the beginning and they grow at a slow rate. Thus, the average grain size and its variance diminishes when $E_N/E_G$ diminishes. In contrast, when $E_N \gg E_G$, during the first stages of crystallization, growth dominates and the nucleation rate increases progressively as crystallization proceeds. Since the time left for growth is less for the nuclei that appear later, the density of small grains will be higher than that of larger grains. So the average grain size and the distribution variance increase with $E_N/E_G$. On the other hand, despite the large variation of $\mathrm{var}_0$, the total variation of the ratio between the exact and the approximate $\mathrm{var}_0$ is very smooth -- from 1.96 to 2.12. A similar behavior has been observed for the 2D-case where this rate evolves from 1.28 at $E_N/E_G=10$ to 1.35 at $E_N/E_G=0$. Hence, the deviations of the approximate value of $\mathrm{var}_0$ from the exact value remain practically constant. This result is due to \textit{approximation I}, which is based on a geometrical approach that is fairly insensitive to the relation between nucleation and growth rates.

\begin{figure}
\includegraphics[width=8.3cm]{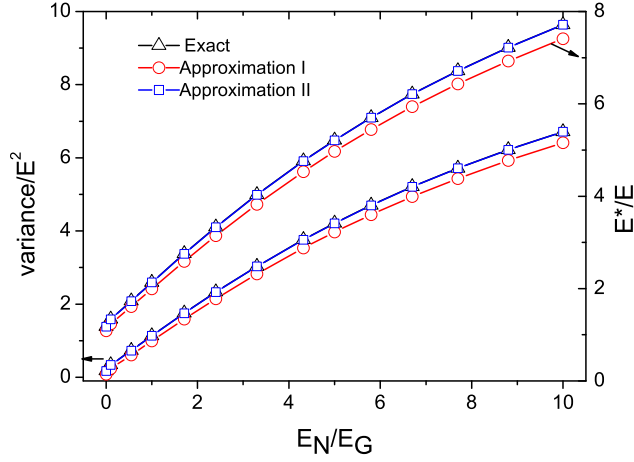}
\caption{\label{fig:evovar} Exact (black triangles), approximate (red circles) and corrected approximate (blue squares) values of the variance and $E^*$ and as a function of the ratio between the nucleation and growth activation energies.}
\end{figure}

Since the ratio between the exact and the approximate $\mathrm{var}_{\tau}$ is nearly constant, we can obtain a significantly more accurate approximate value for $\mathrm{var}_{\tau}$ by simply multiplying it by the corresponding proportionality constant. This constant only depends on the growth dimensionality. We have chosen the values of 2.07 and 1.32 for the 3D and 2D cases, respectively. These values correspond to $E_N/E_G=1$, i.e., they correspond to the isothermal case with $E_G$ and $E_N$ constant in time. With this correction (from now on \textit{approximation II}), the relative error in the calculation of the variance diminishes to less than 2\% and for $E_N/E_G \geq 1$ the relative error is less than 0.2\%. In Fig.~\ref{fig:evovar} we have plotted the exact and the approximate values of the distribution variance and $E^*$ with respect to the ratio $E_N/E_G$. The exact and the approximate value obtained from \textit{approximation II} of the variance and $E^*$ exhibit excellent agreement; the values overlap in such a way that they are nearly indistinguishable. Concerning the values obtained from \textit{approximation I}, it is worth noting that despite the significant error related to the calculation of $\mathrm{var}_0$ [Fig.~\ref{fig:evovartau}] and of $\mathrm{var}_{\tau}$ in general, the inaccuracies in the evaluation of the variance and $E^*$ are significantly smaller. The reason is that both parameters depend exclusively on $E_{\tau}$ and $E$ [Eqs.~(\ref{eq:variantotal}) and (\ref{eq:Eestrella})]. From Fig.~\ref{fig:vartau} it is clear that the error related to $E_{\tau}$ is significantly smaller than the error in the evaluation of $\mathrm{var}_{\tau}$, while the calculation of $E$ is exact.

From now on, when will always use \textit{approximation II} in the calculation of the approximate values of $E^*$ and $E_{\tau}^*$.

\section{Approximate cell size probability density function}

One application of the preceding analysis of the statistical properties of grain size distribution is the derivation of a PDF. If we choose a set of $f_{\tau}(s)$ such that their expected value coincides with the result of Eq.~(\ref{eq:Etau2}) and their variance is equal to the value given by Eq.~(\ref{eq:vartau}), then the variance and expected values of the PDF obtained from Eq.~(\ref{eq:PDFtotal}) will be exact, i.e., the PDF obtained from Eq.~(\ref{eq:PDFtotal}) will have the same variance and expected values as the actual PDF. Indeed, Pineda et al.,\cite{Pineda2004} apply this approach in the case of tessellations generated by random nucleation processes where the growth rate was assumed to be constant. The agreement between their approximate PDF and Monte-Carlo simulations was remarkable. However, they did not notice that the expected value and the variance of their approximate PDF were exact. Concerning the particular choice of the $f_{\tau}(s)$ functions, we will consider two cases: gamma and Gaussian distributions.

\subsection{Gamma distribution}

Given that in a ${\tau}$-distribution the nucleation events are simultaneous and the $\tau$-nuclei are randomly distributed, we can assume that $f_{\tau}(s)$ is similar to the PDF resulting from a process of site saturated nucleation, i.e. that the PDF is that of a Poisson-Voronoi tessellation. As explained in the introduction, the gamma distribution is the exact PDF for the one-dimensional case while it provides a very accurate result for the two- and three-dimensional cases:
\begin{equation}
f_{\tau}(s)=\left( \frac{\nu_{\tau}}{E_{\tau}} \right)^{\nu_{\tau}}
\frac{1}{\Gamma(\nu_{\tau})} s^{\nu_{\tau}-1} \, \text{exp}\left(-\frac{\nu_{\tau} \, s}{E_{\tau}} \right),\nonumber
\end{equation}
Since $E_{\tau}$ is already the expected value, the problem comes down to the determination of the exponent $\nu_{\tau}$. Indeed, $\nu_{\tau}$ can be calculated from the following property of the gamma distributions
\begin{equation}
\nu_{\tau}=\frac{(E_{\tau}) ^2}{\mathrm{var}_{\tau}}
\label{eq:varian}.
\end{equation}

\begin{figure}
\includegraphics[width=8.3cm]{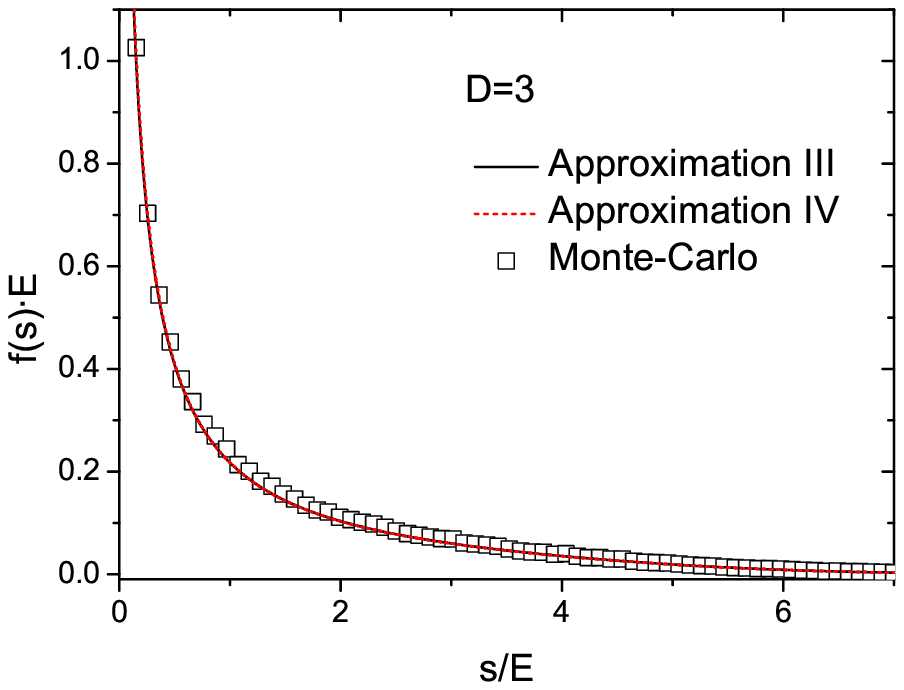}
\caption{\label{fig:Si3D} Grain size distribution for three-dimensional growth and isochronal heating. Comparison between the PDFs obtained from the exact calculation of $E_{\tau}^*$ (black solid line), from the approximate calculation of $E_{\tau}^*$ (red dashed line) and from Monte-Carlo simulation (empty squares).}
\end{figure}

\begin{figure}
\includegraphics[width=8.3cm]{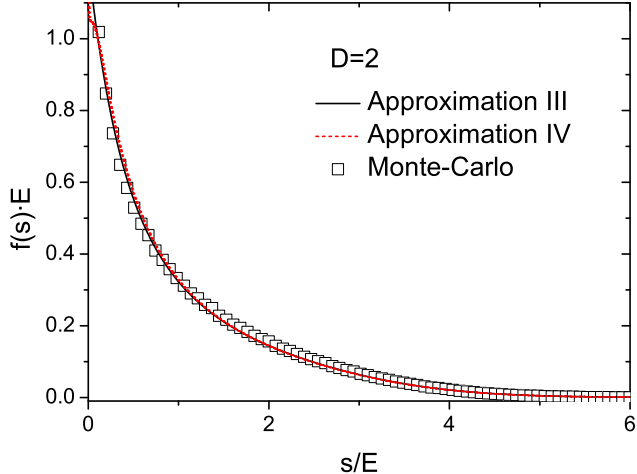}
\caption{\label{fig:Si2D} Grain size distribution for two-dimensional growth and isochronal heating. Comparison between the PDFs obtained from the exact calculation of $E_{\tau}^*$ (black solid line), from the approximate calculation of $E_{\tau}^*$ (red dashed line) and from Monte-Carlo simulation (empty squares).}
\end{figure}

To check the accuracy of the PDF we compare them to some Monte-Carlo simulations. The Monte-Carlo algorithm\cite{Farjas2007} consists in dividing the space into a cubic lattice. Cells are assigned to nuclei randomly. The nucleation time of each nucleus is precisely calculated from the nucleation rate and it is recorded to evaluate the exact evolution of the grain growth. Finally, each cell is assigned to the nucleus that first reaches this cell. The evolution of the grain growth transformation is checked whenever the grain growth is equal to the size of a cell in order to avoid incorrect cell assignation due to shielding effects. In particular we consider the crystallization of amorphous silicon under isochronal heating at 40 K/min (Table~\ref{tab:table1}). The results of the calculations for three- and two-dimensional growth are given in Figs.~(\ref{fig:Si3D}) and (\ref{fig:Si2D}), respectively. We have calculated the PDF using the exact, Eq.~(\ref{eq:PbQfinal}), and approximate, Eq.~(\ref{eq:PbQaprox}), values of $E_{\tau}^*$. We will refer to these PDFs as the \textit{approximate III} and \textit{approximate IV} PDFs, respectively. The validity of the selection of a gamma distribution for the $f_{\tau}(s)$ functions is confirmed by the good agreement between the calculated PDF and the Monte-Carlo simulations. The excellent agreement between the \textit{approximate III} and the \textit{approximate IV} PDFs is also noteworthy  -- the relative difference is less than 0.1\% for $s/E>0.05$.
Therefore, the approximate calculation of $E_{\tau}^*$ is useful to obtain a simple and accurate PDF for a KJMA tessellation. However, the complexity and the computing time required for their evaluation is significantly different. For instance, the calculation of the \textit{approximate III} PDF typically takes more than thirty times the time required for the calculation of the \textit{approximate IV} PDF. The reason for this significant simplification and reduction in computing time is that the \textit{approximate IV} PDF saves us from having to evaluate the inner and more complex integral of Eq.~(\ref{eq:PbQfinal}).

\begin{figure}
\includegraphics[width=8.3cm]{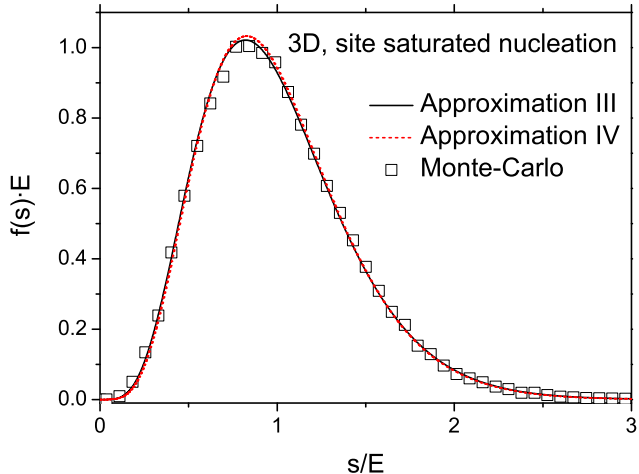}
\caption{\label{fig:Preex3D} Grain size distribution for three-dimensional growth and site saturated nucleation. Comparison between the PDFs obtained from the exact calculation of $E_{\tau}^*$ (black solid line), from the approximate calculation of $E_{\tau}^*$ (red dashed line) and from Monte-Carlo simulation (empty squares).}
\end{figure}

To confirm this last conclusion, we have calculated the PDF for the case of \textit{site saturated nucleation}. This case corresponds to $E_N/E_G=0$ in Figs.~(\ref{fig:evovartau}) and (\ref{fig:evovar}) and is the case that exhibits the greatest discrepancy between the exact and approximate values of the variance and $E^*$. Therefore, it should give the worst agreement between \textit{approximate III} and the \textit{approximate IV} PDFs. As is apparent from Fig.~(\ref{fig:Preex3D}) here again the agreement between the PDFs obtained from the exact calculation of $E_{\tau}^*$, from the approximate calculation of $E_{\tau}^*$ and from Monte-Carlo simulation is excellent. Only small deviations of the \textit{approximate III} from the \textit{approximate IV} PDF are distinguishable for $s\approx E$. Finally, for the one-dimensional case and for site saturated nucleation, both the \textit{approximate III} and \textit{approximate IV} PDFs turn into the exact PDF (see Appendix B).

\subsection{Gaussian distribution}

When analyzing the crystallization morphology, it is often better to use the grain radius distribution instead of the grain size distribution. The grain radius, $r$, of a grain of size $s$, is defined as the radius that will have a $D$-sphere of volume $s$. For instance, for three-dimensional growth
\begin{equation}
 r \equiv \sqrt[3]{\frac{3 s}{4 \pi}}
\label{eq:defr}.
\end{equation}

Given the grain size PDF, the grain radius PDF, $g(r)$, can be easily derived; for three-dimensional growth,
\begin{equation}
 g(r)= 4 \pi r^2 f \left( \frac{4}{3} \pi r^3 \right)
\label{eq:defgr}.
\end{equation}

On the other hand, for site saturated nucleation and three-dimensional growth it has been shown that $g(r)$ is accurately described by a Gaussian distribution\cite{Farjas2007}
\begin{equation}
 g(r)=\frac{1}{\sqrt{2 \pi \sigma}}\text{exp}\left(-\frac{(x-E_g)^2}{2 \sigma^2} \right)
\label{eq:defgauss},
\end{equation}
\noindent where $E_g$ is the expected value of the grain radius PDF and $\sigma$ is the standard deviation
\begin{equation}
 \sigma=\sqrt{\mathrm{var}_g}
\label{eq:defvargauss}.
\end{equation}
\noindent where $\mathrm{var}_g$ is the variance of the grain radius distribution. The fitted parameters were actually $E_g=0.608/n_0^{1/3}$ and $\sigma=0.0883/n_0^{1/3}$, where $n_0$ is the nuclei density.

Conversely, $g(r)$ can be determined by means of Eq.~(\ref{eq:defgr}) where $f(s)$ is a gamma distribution with $\nu=5.581$.\cite{Pineda2004}  Under this approach, it can easily be proved that
\begin{equation}
 E_g=\left( \frac{3}{4 \pi} \frac{E}{\nu}\right)^{1/3}\frac{\Gamma(\nu+1/3)}{\Gamma(\nu)} \nonumber
\end{equation}
\begin{equation}
 \mathrm{var}_g=\left( \frac{3}{4 \pi} \frac{E}{\nu}\right)^{2/3}
\left[\frac{\Gamma(\nu+2/3)}{\Gamma(\nu)} -  \frac{\Gamma(\nu+1/3)^2}{\Gamma(\nu)^2} \right]
\label{eq:relgaussgamma},
\end{equation}
\noindent where $E=1/n_0$ is the expected value of the grain size distribution (see Appendix B). By substituting $\nu=5.581$ into the previous relations we obtain $E_g=0.608/n_0^{1/3}$, $\mathrm{var}_g=7.65\, 10^{-3}/n_0^{2/3}$ and $\sigma=0.0874/n_0^{1/3}$, which are in good agreement with the parameters obtained from the Gaussian fit.

\begin{figure}
\includegraphics[width=8.3cm]{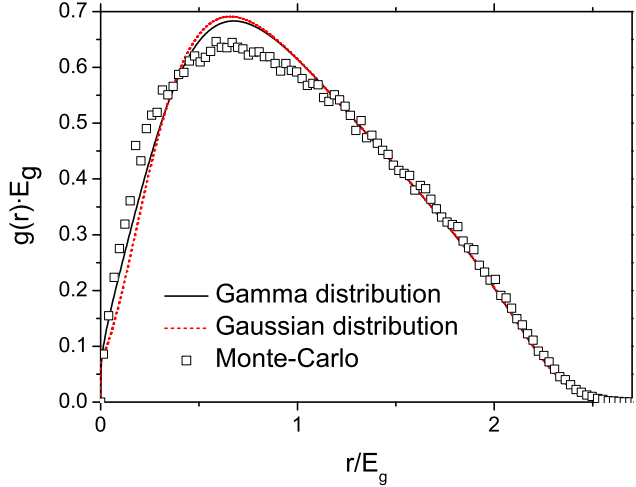}
\caption{\label{fig:FigGauss} Grain radius distribution for three-dimensional growth and isochronal heating. Comparison between the PDF obtained from Eq.~(\ref{eq:defgr}) (black solid line), Eq.~(\ref{eq:PDFtotalR}) (red dashed line) and  Monte-Carlo simulation (empty squares).}
\end{figure}

Therefore, for a KJMA tessellation we can obtain $g(r)$ as the superposition of $g_{\tau}(r)$ PDFs. In contrast with the previous subsection, we will now assume that $g_{\tau}(r)$ are Gaussian distributions:
\begin{equation}
 g(r)=\frac{\int_0^{\infty}I_a(\tau) g_{\tau}(r) d \tau}{\int_0^{\infty}I_a(\tau) d \tau}
\label{eq:PDFtotalR}.
\end{equation}
To evaluate the $g_{\tau}(r)$ PDFs we need to know their expected value, $E_{g,\tau}$, and their variance, $\mathrm{var}_{g,\tau}$. These parameters can be easily derived from the statistical parameters of the ${\tau}$-distributions, $E_{\tau}$ and $\mathrm{var}_{\tau}$, by means of Eq.~(\ref{eq:relgaussgamma}). In Fig.~(\ref{fig:FigGauss}) we have plotted the grain radius distribution obtained, from the grain size distribution [Eq.~(\ref{eq:defgr})] (where $f(s)$ is the superposition of gamma distributions), as a direct superposition of $g_{\tau}(r)$ Gaussian PDFs [Eq.~(\ref{eq:PDFtotalR})] and from Monte-Carlo simulations. This distribution corresponds to the crystallization of amorphous silicon under isochronal heating at 40 K/min (Table~\ref{tab:table1}). The good agreement between the PDF calculated from Eqs.~(\ref{eq:defgr}) and (\ref{eq:PDFtotalR}) confirms that the $g_{\tau}(r)$ are correctly described by a Gaussian distribution. Finally, both approaches show good agreement with Monte-Carlo simulations, i.e., both approaches are useful for describing the grain radius PDF.

\section{Conclusions}

This paper deals with a subject of interest in many scientific areas, namely the cell-size distribution of space tessellations that emerge from first-order phase transformations ruled by nucleation and growth of the new stable phase. No restrictions are imposed on the time dependency of the nucleation and growth rates, and the validity of our results is limited to transformations that obey the premises of the Kolmogorov-Johnson-Mehl-Avrami model. We have derived some important statistical properties such as the expected value and the variance. The approach used is an extension of the work of Gilbert.\cite{Gilbert1962} Additionally, we have developed a significantly simpler relation for the calculation of the variance. The discrepancies between the exact and approximate variances are less than 2\%.

Like Pineda et al.,\cite{Pineda2004} we have derived an approximate grain size PDF as the superposition of gamma distributions. We have proved that the expected value and variance derived from this approximate grain size PDF are exact. Moreover, we have checked its accuracy against Monte-Carlo simulations for a system undergoing a crystallization under isochronal heating conditions. The results show a remarkably good agreement between the approximate PDF and the Monte-Carlo simulations. Finally, we have shown that the grain radius PDF can be expressed as the superposition of Gaussian distributions.

\appendix

\section{One-dimensional PDF}

For the one dimensional case, the extended transformed fraction is
\begin{equation}
 X_{ex}(t)=2 \int_0^t I(\tau) \left[ \int_{\tau}^t G(z) d z \right] d\tau
\label{eq:Xex1D},
\end{equation}
\noindent and the expected value $E_{\tau}$ becomes
\begin{equation}
E_{\tau}=\frac{2}{1-X(\tau)}\int_{\tau}^{\infty} \left[ 1-X(z) \right] \, G(z) d z
\label{eq:Etau1D}.
\end{equation}
With regard to the calculation of $P_{\tau}^*(b)$, we have split the entire space into three regions. The first corresponds to $Q$ located at the left side of $O$, in this case $r_P=r_O+b$ and
\begin{equation}
 P_{\tau}^*(b,Q)=\exp[-X_{ex}(t_P)]
\label{eq:regio1}.
\end{equation}
The second region corresponds to $Q$ located between $O$ and $P$, then $r_P+r_O=b$ and
\begin{equation}
 P_{\tau}^*(b,Q)=\exp[-X_{ex}(t_O)-X_{ex}(t_P)+X_{ex}(\tau)]
\label{eq:regio2}.
\end{equation}
And the third region corresponds to $Q$ located at the right side of $P$, $r_O=r_P+b$ and
\begin{equation}
 P_{\tau}^*(b,Q)=\exp[-X_{ex}(t_O)]
\label{eq:regio3}.
\end{equation}
Finally, combining Eqs.~(\ref{eq:regio1}), (\ref{eq:regio2}) and (\ref{eq:regio3}) with (\ref{eq:PbEst}) we obtain
\begin{eqnarray}
 P_{\tau}^*(b) & = & I(\tau) \Big\{ 2 \int_b^\infty [1-X(t_O)] d r_O +\nonumber \\
	     &   &  + \frac{1}{1-X(\tau)} \int_0^b [1-X(t_O)] [1-X(t_P)] d r_O\Big\}\nonumber,
\end{eqnarray}
\begin{equation}
 r_O=\int_{\tau}^{t_O} G(z) dz \nonumber,
\end{equation}
\begin{equation}
 b-r_O=\int_{\tau}^{t_P} G(z) dz
\label{eq:Pb1D}.
\end{equation}
Once $P_{\tau}^*(b)$ is known, Eq.~(\ref{eq:Eestrellatau}) combined with Eq.~(\ref{eq:vartau}) delivers $\mathrm{var}_{\tau}$.

\section{Site saturated nucleation}
When nucleation is completed prior to crystal growth, the nucleation rate can be approximated to ${I(t)\approx n_{0}\mathit{\delta}(t)}$ where $n_0$ is the density of nuclei and $\delta$ is the Dirac delta function. In this case, the extended transformed fraction becomes
\begin{equation}
  X_{ex}(t)= n_0 g_D \left[ \int_{0}^t  G(z) d z \right]^D
\label{eq:XexNP}.
\end{equation}
Then
\begin{equation}
  I_a(\tau)= n_0[1-X(\tau)]\delta(\tau)
\end{equation}
\noindent and the PDF, Eq.~(\ref{eq:PDFtotal}), is reduced to
\begin{equation}
f(s)=\frac{\int_0^{\infty}n_0[1-X(\tau)]\delta(\tau) f_{\tau}(s) d \tau}
{\int_0^{\infty}n_0[1-X(\tau)]\delta(\tau) d \tau}=f_0(s)
\label{eq:PDFtotalNP}.
\end{equation}
Therefore, in this case we only need to calculate $E_0$ and $\mathrm{var}_0$. Concerning $E_0$, Eq.~(\ref{eq:Etau2}), it can be easily proved that it is simply
\begin{equation}
E_0=\frac{1}{n_0}\int_0^\infty \exp[-X_{ex}] d X_{ex}=1/n_0
\label{eq:EtauNP}.
\end{equation}
Indeed, according to Eq.~(\ref{eq:PDFtotalNP}) $E_0=E$, and the expected value of $E$ is  $1/n_0$ [Eq.~(\ref{eq:EsExt})]. Moreover, the fraction of space occupied by a $\tau$-crystal, $X_{\tau}$, is reduced to  $\delta(\tau)$, as expected.

With respect to $\mathrm{var}_0$, its value is determined by $E_0^*$ and $E_0$ by means of Eq.~(\ref{eq:vartau}). $E_0^*$ is given by Eq.~(\ref{eq:Eestrellatau}). We therefore need to evaluate $P_0^*(b)$, the value of which depends on which relation for $P_0^*(b,Q)$ we use: the exact one Eq.~(\ref{eq:PbQfinal}) or the approximate one Eq.~(\ref{eq:PbQaprox}). We will evaluate $P_0^*(b)$ for the one-dimensional case because in this case the approximate solution coincides with the exact one. Specifically, substitution of Eq.~(\ref{eq:XexNP}) into Eq.~(\ref{eq:Pb1D}) leads to
\begin{eqnarray}
P_0^*(b) & = & n_0 \delta(\tau) \Big[2 \int_b^\infty e^{-2 n_0 r_O} d r_O+\int_0^b e^{-2 n_0 b} d r_O \Big]=\nonumber \\
       & = & n_0 \delta(\tau) e^{-2 n_0 b} \left( 1/n_0+b\right)
\label{eq:Pb1DNP}.
\end{eqnarray}
Then, from Eqs.~(\ref{eq:Eestrellatau}) and (\ref{eq:Pb1DNP})
\begin{equation}
 E_0^*=\frac{3}{2} \frac{1}{n_0}
\label{eq:Eestr1DNP}.
\end{equation}
And finally, from Eqs.~(\ref{eq:vartau}) and (\ref{eq:Eestr1DNP}),
\begin{equation}
 \mathrm{var}_0 = \left( 2 n_0 \right)^{-1}
\label{eq:nu1DNP}.
\end{equation}
If we choose a gamma distribution for the calculation of the cell-size PDF, from Eq.~(\ref{eq:varian}) we obtain $\nu_0=2$. Thus for site-saturated nucleation and one-dimensional growth we obtain a gamma distribution with $\nu=2$ and $E=1/n_0$ which agrees with the exact solution.\cite{Meijering1953,Axe1986}

\begin{acknowledgments}
This work has been supported by the Spanish Programa Nacional de Materiales under contract number MAT2006-11144 and by the Generalitat de Catalunya under contract number 2005SGR-00666.
\end{acknowledgments}

\bibliography{Referencies}

\end{document}